\documentstyle[seceq,preprint]{ptptex}

\preprintnumber{TOYAMA-90}
\title{
Massless Limits of Massive Tensor Fields
}

\author{
Shinji {\sc Hamamoto}
\footnote{E-mail address: hamamoto@sci.toyama-u.ac.jp}
}

\inst{
Department of Physics, Toyama University, Toyama 930
}

\recdate{
\today
}

\abst{
In order to construct a massive tensor theory 
with a smooth massless limit, 
we apply two kinds of gauge-fixing procedures, 
Nakanishi's one and the BRS one, 
to two models of massive tensor field. 
The first is of the Fierz-Pauli (FP) type, 
which describes a pure massive tensor field; 
the other is of the additional-scalar-ghost (ASG) type, 
which includes a scalar ghost in addition to an ordinary tensor field. 
It is shown that 
Nakanishi's procedure can eliminate massless singularities 
in both two models, 
while the BRS procedure regularizes the ASG model only.
The BRS-regularized ASG model is most promising 
in constructing a complete nonlinear theory.
}

\begin{document}

\maketitle

\section{Introduction}
In order to obtain a satisfactory formulation of the infrared problem 
in quantum gravity, 
we re-examine smooth massless limits of massive tensor field theories. 

Two models are studied in the present paper: 
the first is of the Fierz-Pauli (FP) type; 
the other is of the additional-scalar-ghost (ASG) type.
The FP model has been adopted as a standard model of massive tensor 
field 
because it describes a pure massive tensor 
with five degrees of freedom. 
Two-point functions of this model, however, 
take form quite different from the corresponding ones in the massless case. 
On the other hand, although the ASG model includes a scalar ghost 
in addition to an ordinary tensor field, 
there are some similarities between two-point functions of the ASG model 
and those in the massless case.\cite{rf:1} 

Kimura\cite{rf:2} investigated a massless tensor field 
in general covariant gauge, 
proposing a model with ASG-type mass term as a good candidate 
for massive theory with a smooth massless limit. 
Fronsdal and Heidenreich\cite{rf:3} succeeded in regularizing the FP model. 
They subtracted every massless singularity 
from the whole set of two-point functions 
by introducing two kinds of auxiliary fields of spin-1 and 0. 
Because the Lagrangian obtained is not so simple, however, 
it seems unsuitable for constructing a complete nonlinear theory. 

In the present paper we apply to the two models 
two kinds of gauge-fixing procedures: 
Nakanishi's procedure and the BRS one. 
Nakanishi showed that
simple addition of a gauge-fixing term 
to the free Lagrangian for an Abelian massive vector field can 
regularize massless singularities in the original theory.\cite{rf:4} 
\ Applying this procedure to the case of tensor field, 
we find that both of the two models become free 
from massless singularities. 
The BRS gauge-fixing procedure has been recognized to have 
wide applicability.\cite{rf:5} 
\ It is shown that 
massless singularities in the ASG model 
are in fact regularized by this procedure. 
For the FP model, however, this procedure does not work: 
there still remain massless singularities, 
though weaker than before the application of this procedure. 

However, this is the story of a linearized world. 
Our main, not present but future, purpose is 
to construct a complete nonlinear theory of massive 
tensor field.
For this purpose
it is desirable to have linearized theories 
with higher symmetry properties. 
From this point of view the BRS procedure is more suitable than 
Nakanishi's. 
This is because the former installs BRS symmetry in the theory, 
while the latter does not implement any symmetry property. 
That means the BRS-regularized ASG model seems most promising 
for our purpose. 

In \S 2, we review the case of Abelian vector field. 
This is to see how Nakanishi's and the BRS procedures work 
for regularizing massless singularities in the original massive theory. 
In \S 3, two models are presented for massive tensor field. 
Nakanishi's gauge-fixing procedure is applied to them in \S 4, 
while the BRS procedure applied in \S 5. 
Section 6 is devoted to summary and discussion.

\section{Massive Vector Fields}

\subsection{Massless vector}
We begin with a free Abelian massless vector field.
The Lagrangian with the usual gauge-fixing term is 
\footnote{The metric used in the present paper is 
$\eta^{\mu\nu} = ( -1, +1, +1, +1 )$.}
\begin{equation}
L = \mbox{}-\frac{1}{4}F_{\mu\nu}F^{\mu\nu} 
+ b \left(\partial^{\mu}A_{\mu} + \frac{\alpha}{2}b\right) ,
\end{equation}
where $b$ is the Nakanishi-Lautrup (NL) field 
and $\alpha$ is the gauge parameter. 
Field equations are
\begin{eqnarray}
\Box A_{\mu} - ( 1 - \alpha ) \partial_{\mu}b & = & 0 , \\
\partial^{\mu}A_{\mu} + \alpha b & = & 0 , \\
\Box b & = & 0 .
\end{eqnarray}
We also have
\begin{eqnarray}
\Box \partial^{\mu}A_{\mu} & = & 0 , \\
\Box^{2}A_{\mu} & = & 0 .
\end{eqnarray}
Two-point functions are
\begin{eqnarray}
\langle A^{\mu}(x)A^{\nu}(y)\rangle & = & 
\frac{1}{\Box}\left[\eta^{\mu\nu} - 
(1 - \alpha )\frac{\partial^{\mu}\partial^{\nu}}{\Box}\right]
\delta ( x - y ) ,\\
\langle A^{\mu}(x)b(y)\rangle & = & 
\frac{\partial^{\mu}}{\Box}\delta (x - y ) ,\\
\langle b(x)b(y)\rangle & = & 0 .
\end{eqnarray}

\subsection{Massive vector}
In this case the Lagrangian is given by 
\begin{equation}
L = \mbox{}-\frac{1}{4}F_{\mu\nu}F^{\mu\nu} 
- \frac{m^{2}}{2}A_{\mu}A^{\mu} .
\label{eq:210}
\end{equation}
Field equations are 
\begin{eqnarray}
\left(\Box - m^{2}\right)A_{\mu} & = & 0 , \\
\partial^{\mu}A_{\mu} & = & 0 .
\end{eqnarray}
Two-point functions are
\begin{equation}
\langle A^{\mu}(x)A^{\nu}(y)\rangle = 
\frac{1}{\Box - m^{2}}\left[\eta^{\mu\nu} - 
\frac{\partial^{\mu}\partial^{\nu}}{m^{2}}\right]\delta ( x - y ) ,
\end{equation}
which develop massless singularities in the limit of 
$m = 0$.

\subsection{Nakanishi's gauge-fixing procedure}
Following Nakanishi,\cite{rf:4} we add to the Lagrangian (\ref{eq:210}) 
the same gauge-fixing term as in the massless case: 
\begin{equation}
L = \mbox{}-\frac{1}{4}F_{\mu\nu}F^{\mu\nu} 
- \frac{m^{2}}{2}A_{\mu}A^{\mu}
+ b\left(\partial^{\mu}A_{\mu} + \frac{\alpha}{2}b\right) .
\end{equation}
This yields the following field equations: 
\begin{eqnarray}
\left(\Box - m^{2}\right)A_{\mu} - ( 1 - \alpha ) \partial_{\mu}b & = & 0 , \\
\partial^{\mu}A_{\mu} + \alpha b & = & 0 , \\
\left(\Box - \alpha m^{2}\right) b & = & 0 ,
\end{eqnarray}
and
\begin{eqnarray}
\left(\Box - \alpha m^{2}\right)\partial^{\mu}A_{\mu} & = & 0 , \\
\left(\Box - \alpha m^{2}\right)\left(\Box - m^{2}\right)A_{\mu} & = & 0 .
\end{eqnarray}
Two-point functions in this case are 
\begin{eqnarray}
\langle A^{\mu}(x)A^{\nu}(y)\rangle & = & 
\frac{1}{\Box - m^{2}}\left[\eta^{\mu\nu} - 
(1 - \alpha )\frac{\partial^{\mu}\partial^{\nu}}{\Box - \alpha m^{2}}\right] 
\delta ( x - y ) ,\\
\langle A^{\mu}(x)b(y)\rangle & = & 
\frac{\partial^{\mu}}{\Box - \alpha m^{2}}\delta ( x - y ) ,\\
\langle b(x)b(y)\rangle & = & 
\mbox{}-\frac{m^{2}}{\Box - \alpha m^{2}}\delta ( x - y ) ,
\end{eqnarray}
which show that the massless singularities in the original massive theory 
have been regularized by this procedure.

\subsection{BRS gauge-fixing procedure}
General consideration of this procedure 
has been developed 
by Izawa.\cite{rf:5}  \ For the massive vector case, 
starting from the usual Lagrangian (\ref{eq:210})
\begin{equation}
L_{A} = \mbox{}-\frac{1}{4}F_{\mu\nu}F^{\mu\nu} 
- \frac{m^{2}}{2}A_{\mu}A^{\mu} ,
\label{eq:223}
\end{equation}
we intend to carry out a field tranformation 
$A_{\mu} \rightarrow ( A_{\mu}', \theta )$ such that
\begin{eqnarray}
A_{\mu} & = & A_{\mu}' - \frac{1}{m}\partial_{\mu}\theta , \\
\partial^{\mu}A_{\mu}' & = & 0 .
\end{eqnarray}
Since the Lagrangian (\ref{eq:223}) is independent of 
the new variables $( A_{\mu}', \theta )$, 
it is invariant under the BRS transformation 
\begin{equation}
\left\{
\begin{array}{lcl}
\delta A_{\mu}' = c_{\mu} , & & \delta \bar{c}_{\mu} = ib_{\mu} , \\
\delta \theta = mc ,        & & \delta \bar{c} = ib ,
\end{array}
\right.
\end{equation}
where the Faddeev-Popov (FP) ghosts 
$(c_{\mu}, c)$ and $(\bar{c}_{\mu},\bar{c})$ 
as well as the NL fields $(b_{\mu}, b)$ have been introduced. 
To relate the old and new sets of variables, 
we add to the Lagrangian (\ref{eq:223}) the following BRS term: 
\begin{eqnarray}
L_{\rm B} & = & 
\mbox{}-i\delta\left[ 
\bar{c}^{\mu}\left(A_{\mu} - A_{\mu}' + \frac{1}{m}\partial_{\mu}\theta\right) 
+ \bar{c}\left(\partial^{\mu}A_{\mu}' + \frac{\alpha}{2}b\right)\right] 
\nonumber \\
& = & \,
b^{\mu}\left(A_{\mu} - A_{\mu}' + \frac{1}{m}\partial_{\mu}\theta\right) 
+ b\left(\partial^{\mu}A_{\mu}' + \frac{\alpha}{2}b\right) \nonumber \\
& & \makebox[15mm]{}
-i\left(\bar{c}^{\mu} + \partial^{\mu}\bar{c}\right) 
\left(c_{\mu} - \partial_{\mu}c\right) 
+ i\bar{c}\Box c .
\end{eqnarray}
The path integral is given as
\begin{equation}
Z = \int {\cal D}A_{\mu}{\cal D}A_{\mu}'{\cal D}\theta
{\cal D}b_{\mu}{\cal D}c_{\mu}{\cal D}\bar{c}_{\mu}
{\cal D}b{\cal D}c{\cal D}\bar{c}
\exp i\int d^{4}x\left[L_{A} + L_{\rm B}\right] .
\end{equation}
Integrating over the variables 
$(b_{\mu}, A_{\mu}, c_{\mu}, \bar{c}_{\mu})$ 
and overwriting $A_{\mu}$ on $A_{\mu}'$, we obtain 
\begin{equation}
Z = \int {\cal D}A_{\mu}{\cal D}\theta{\cal D}b{\cal D}c{\cal D}\bar{c}
\exp i\int d^{4}x\, L_{\rm T} ,
\end{equation}
where
\begin{equation}
L_{\rm T} = \mbox{}-\frac{1}{4}F_{\mu\nu}F^{\mu\nu} 
- \frac{m^{2}}{2}\left(A_{\mu} - \frac{1}{m}\partial_{\mu}\theta\right)^{2}
+ b\left(\partial^{\mu}A_{\mu} + \frac{\alpha}{2}b\right) 
+ i\bar{c}\Box c .
\label{eq:230}
\end{equation}
This Lagrangian is invariant under the following BRS transformation: 
\begin{equation}
\begin{array}{lclcl}
\delta A_{\mu} = \partial_{\mu}c , & & 
\delta\theta = mc ,                & & 
\delta\bar{c} = ib . 
\end{array}
\end{equation}
Since our model is Abelian, the FP ghosts $(c, \bar{c})$ 
decouple from any other field in the Lagrangian (\ref{eq:230}). 
When discussing field equations and two-point functions, therefore, 
we can neglect the last term in $L_{\rm T}$. 
What we have obtained is nothing but the Stueckelberg Lagrangian. 
Field equations are 
\begin{eqnarray}
\left(\Box - m^{2}\right)A_{\mu} - (1 - \alpha )\partial_{\mu}b 
+ m\partial_{\mu}\theta = 0 , & & \\
\partial^{\mu}A_{\mu} + \alpha b = 0 , & & \\
\Box b = 0 , & & \\
\Box\theta + \alpha mb = 0 , & & 
\end{eqnarray}
and 
\begin{eqnarray}
\Box\partial^{\mu}A_{\mu} = 0 , & & \\
\Box^{2}\theta = 0 , & & \\
\Box^{2}\left(\Box - m^{2}\right)A_{\mu} = 0 . & & 
\end{eqnarray}
Two-point functions are 
\begin{eqnarray}
\langle A^{\mu}(x)A^{\nu}(y)\rangle & = & 
\frac{1}{\Box - m^{2}}\left[\eta^{\mu\nu} 
- (1 - \alpha )\frac{\partial^{\mu}\partial^{\nu}}{\Box}
- \alpha m^{2}\frac{\partial^{\mu}\partial^{\nu}}{\Box^{2}}
\right]
\delta ( x - y ) ,\nonumber \\
& & \\
\langle A^{\mu}(x)b(y)\rangle & = & 
\frac{\partial^{\mu}}{\Box}\delta (x - y ) ,\\
\langle A^{\mu}(x)\theta (y)\rangle & = & \mbox{}
- \alpha m\frac{\partial^{\mu}}{\Box^{2}}\delta (x-y) ,\\
\langle b(x)b(y)\rangle & = & 0 ,\\
\langle b(x)\theta (y)\rangle & = & 
m\frac{1}{\Box}\delta (x-y) , \\
\langle\theta (x)\theta (y)\rangle & = & 
\left(\frac{1}{\Box} - \alpha m^{2}\frac{1}{\Box^{2}}\right)\delta (x-y) .
\end{eqnarray}
They are in fact singular-free in the massless limit.
The field $\theta$ becomes redundant in this limit 
and the theory smoothly reduces to the usual massless theory.

\section{Massive Tensor Fields}

\subsection{Massless tensor}
The Lagrangian with a gauge-fixing term is given by 
\begin{eqnarray}
L & = & \mbox{}
- \frac{1}{2}\left( 
\partial_{\lambda}h_{\mu\nu}\partial^{\lambda}h^{\mu\nu} 
- \partial_{\lambda}h\partial^{\lambda}h\right) 
+ \partial_{\lambda}h_{\mu\nu}\partial^{\nu}h^{\mu\lambda} 
- \partial_{\mu}h^{\mu\nu}\partial_{\nu}h \nonumber \\
& & \makebox[25mm]{}
+ b^{\mu}\left(
\partial^{\nu}h_{\mu\nu} - \frac{1}{2}\partial_{\mu}h 
+ \frac{\alpha}{2}b_{\mu}\right) \nonumber \\
& = & \,
\frac{1}{2}h^{\mu\nu}\Lambda_{\mu\nu ,\rho\sigma}h^{\rho\sigma} 
+ b^{\mu}\left(
\partial^{\nu}h_{\mu\nu} - \frac{1}{2}\partial_{\mu}h 
+ \frac{\alpha}{2}b_{\mu}\right) ,
\end{eqnarray}
where $h = h^{\mu}_{\makebox[1mm]{}\mu}$ and 
\begin{eqnarray}
\Lambda_{\mu\nu ,\rho\sigma} & = & 
\left(\eta_{\mu\rho}\eta_{\nu\sigma} - \eta_{\mu\nu}\eta_{\rho\sigma}
\right)\Box 
\nonumber \\
& & \mbox{}- \left(\eta_{\mu\rho}\partial_{\nu}\partial_{\sigma} 
+ \eta_{\nu\sigma}\partial_{\mu}\partial_{\rho}\right)
+ \left(\eta_{\rho\sigma}\partial_{\mu}\partial_{\nu} 
+ \eta_{\mu\nu}\partial_{\rho}\partial_{\sigma}\right) .
\end{eqnarray}
Field equations reduce to 
\begin{eqnarray}
\Box h_{\mu\nu} - \frac{1}{2}(1-2\alpha )
\left(\partial_{\mu}b_{\nu} + \partial_{\nu}b_{\mu}\right) 
& = & 0 , \\
\partial^{\nu}h_{\mu\nu} - \frac{1}{2}\partial_{\mu}h + \alpha b_{\mu} 
& = & 0 , \\
\Box b_{\mu} & = & 0 .
\end{eqnarray}
We also have 
\begin{eqnarray}
\Box\left(\partial^{\nu}h_{\mu\nu} - \frac{1}{2}\partial_{\mu}h\right) 
& = & 0 , \\
\Box^{2}h_{\mu\nu} & = & 0 .
\end{eqnarray}
Two-point functions are calculated as
\footnote{Here and hereafter space-time coordinates are omitted 
in the field variables as well as in the $\delta$-functions.
} 
\begin{eqnarray}
\langle h^{\mu\nu}h^{\rho\sigma}\rangle & = & 
\frac{1}{\Box}\left\{
\frac{1}{2}\left(
\eta^{\mu\rho}\eta^{\nu\sigma} 
+ \eta^{\mu\sigma}\eta^{\nu\rho} 
- \eta^{\mu\nu}\eta^{\rho\sigma}\right) \right. \nonumber \\
& & \left. \mbox{}
- \frac{1}{2}(1-2\alpha )\frac{1}{\Box}\left(
\eta^{\mu\rho}\partial^{\nu}\partial^{\sigma}
+ \eta^{\mu\sigma}\partial^{\nu}\partial^{\rho}
+ \eta^{\nu\rho}\partial^{\mu}\partial^{\sigma}
+ \eta^{\nu\sigma}\partial^{\mu}\partial^{\rho}\right)\right\}\delta , 
\nonumber \\
& & \label{eq:308} \\
\langle h^{\mu\nu}b^{\rho}\rangle & = & 
\frac{1}{\Box}\left( 
\eta^{\mu\rho}\partial^{\nu} 
+ \eta^{\nu\rho}\partial^{\mu}\right)\delta , \\
\langle b^{\mu}b^{\rho}\rangle & = & 0 .
\end{eqnarray}

\subsection{Massive tensor of the FP type}
The FP-type Lagrangian for a massive tensor field is given by
\begin{equation}
L = \frac{1}{2}h^{\mu\nu}\Lambda_{\mu\nu ,\rho\sigma}h^{\rho\sigma} 
- \frac{m^{2}}{2}\left( h^{\mu\nu}h_{\mu\nu} - h^{2}\right) .
\label{eq:311}
\end{equation}
The set of field equations
\begin{eqnarray}
\left(\Box - m^{2}\right) h_{\mu\nu} & = & 0 , \\
\partial^{\nu}h_{\mu\nu} & = & 0 , \\
h & = & 0 \label{eq:314}
\end{eqnarray}
shows that the Lagrangian (\ref{eq:311}) purely describes a 
massive tensor field 
with five degrees of freedom.
This is the reason why this type of model has been taken as a standard one. 
However, some undesirable properties are owned by the two-point functions 
\begin{eqnarray}
\langle h^{\mu\nu}h^{\rho\sigma}\rangle & = & 
\frac{1}{\Box - m^{2}}\left\{
\frac{1}{2}\left(
\eta^{\mu\rho}\eta^{\nu\sigma} 
+ \eta^{\mu\sigma}\eta^{\nu\rho} 
- \eta^{\mu\nu}\eta^{\rho\sigma}\right) \right. \nonumber \\
& & \makebox[15mm]{}
- \frac{1}{2m^{2}}\left(
\eta^{\mu\rho}\partial^{\nu}\partial^{\sigma}
+ \eta^{\mu\sigma}\partial^{\nu}\partial^{\rho}
+ \eta^{\nu\rho}\partial^{\mu}\partial^{\sigma}
+ \eta^{\nu\sigma}\partial^{\mu}\partial^{\rho}\right) \nonumber \\
& & \makebox[14mm]{}\left.\mbox{}
+ \frac{2}{3}\left(
\frac{1}{2}\eta^{\mu\nu} 
+ \frac{\partial^{\mu}\partial^{\nu}}{m^{2}}\right)\left(
\frac{1}{2}\eta^{\rho\sigma} 
+ \frac{\partial^{\rho}\partial^{\sigma}}{m^{2}}\right)
\right\}\delta . 
\label{eq:315}
\end{eqnarray}
The first and second terms on the right hand side of this expression 
have their own correspondents in the expressin (\ref{eq:308}).
The massless singularities in the second term are the same 
as encountered in the case of vector field. 
The third term, however, 
develops higher massless singularities than the second term. 
Moreover, that term does not find its own correspondent in the 
expression (\ref{eq:308}). 
These points make this model difficult to regularize.

\subsection{Massive tensor of the ASG type}
We adopt a mass term 
slightly different from the FP-type one: 
\begin{equation}
L = \frac{1}{2}h^{\mu\nu}\Lambda_{\mu\nu ,\rho\sigma}h^{\rho\sigma} 
- \frac{m^{2}}{2}\left( h^{\mu\nu}h_{\mu\nu} - \frac{1}{2}h^{2}\right) .
\end{equation}
In this case, a field equation corressponding to 
(\ref{eq:314}) does not hold. 
We have only
\begin{eqnarray}
\left(\Box - m^{2}\right) h_{\mu\nu} & = & 0 , \\
\partial^{\nu}h_{\mu\nu} 
- \frac{1}{2}\partial_{\mu}h & = & 0 .
\end{eqnarray}
Therefore, this model describes not only an ordinary tensor field 
but also an auxiliary scalar field. 
Two-point functions in this model are given as
\begin{subeqnarray}
\langle h^{\mu\nu}h^{\rho\sigma}\rangle & = & 
\frac{1}{\Box - m^{2}}\left\{
\frac{1}{2}\left(
\eta^{\mu\rho}\eta^{\nu\sigma} 
+ \eta^{\mu\sigma}\eta^{\nu\rho} 
- \eta^{\mu\nu}\eta^{\rho\sigma}\right) \right. \nonumber \\
& & \makebox[10mm]{}\left.\mbox{}
- \frac{1}{2m^{2}}\left(
\eta^{\mu\rho}\partial^{\nu}\partial^{\sigma}
+ \eta^{\mu\sigma}\partial^{\nu}\partial^{\rho}
+ \eta^{\nu\rho}\partial^{\mu}\partial^{\sigma}
+ \eta^{\nu\sigma}\partial^{\mu}\partial^{\rho}\right)\right\}\delta 
\nonumber \\
& & \slabel{eq:319a} \\
& = & 
\frac{1}{\Box - m^{2}}\left\{
\frac{1}{2}\left(
\eta^{\mu\rho}\eta^{\nu\sigma} 
+ \eta^{\mu\sigma}\eta^{\nu\rho} 
- \eta^{\mu\nu}\eta^{\rho\sigma}\right) \right. \nonumber \\
& & \makebox[15mm]{}
- \frac{1}{2m^{2}}\left(
\eta^{\mu\rho}\partial^{\nu}\partial^{\sigma}
+ \eta^{\mu\sigma}\partial^{\nu}\partial^{\rho}
+ \eta^{\nu\rho}\partial^{\mu}\partial^{\sigma}
+ \eta^{\nu\sigma}\partial^{\mu}\partial^{\rho}\right) \nonumber \\
& & \makebox[14mm]{}\left.\mbox{}
+ \frac{2}{3}\left(
\frac{1}{2}\eta^{\mu\nu} 
+ \frac{\partial^{\mu}\partial^{\nu}}{m^{2}}\right)\left(
\frac{1}{2}\eta^{\rho\sigma} 
+ \frac{\partial^{\rho}\partial^{\sigma}}{m^{2}}\right)
\right\}\delta \nonumber \\
& & \mbox{}
- \frac{1}{\Box - m^{2}}\frac{2}{3}\left(
\frac{1}{2}\eta^{\mu\nu} 
+ \frac{\partial^{\mu}\partial^{\nu}}{m^{2}}\right)\left(
\frac{1}{2}\eta^{\rho\sigma} 
+ \frac{\partial^{\rho}\partial^{\sigma}}{m^{2}}\right)\delta .
\slabel{eq:319b}
\end{subeqnarray}
Contrary to the FP model, the expression (\ref{eq:319a}) does not have 
a term like the third one on the right hand side of Eq.(\ref{eq:315}).
This fact simplifies the procedures 
for constructing massless-regular theories.
The expression (\ref{eq:319b}), however, shows that this model includes 
an additional scalar field with negative metric as well as an ordinary 
tensor field.

\section{Nakanishi's Gauge-Fixing Procedure}

\subsection{FP model}
We supplement the FP Lagrangian (\ref{eq:311}) by the same gauge-fixing term 
as in the massless case: 
\begin{equation}
L = \frac{1}{2}h^{\mu\nu}\Lambda_{\mu\nu ,\rho\sigma}h^{\rho\sigma} 
- \frac{m^{2}}{2}\left( h^{\mu\nu}h_{\mu\nu} - h^{2}\right) 
+ b^{\mu}\left(
\partial^{\nu}h_{\mu\nu} - \frac{1}{2}\partial_{\mu}h 
+ \frac{\alpha}{2}b_{\mu}\right) .
\end{equation}
Field equations obtained are
\begin{eqnarray}
\left(\Box - m^{2}\right) h_{\mu\nu} 
- \frac{1}{2}\eta_{\mu\nu}m^{2}h 
- \frac{1}{2}(1-2\alpha )
\left(\partial_{\mu}b_{\nu} + \partial_{\nu}b_{\mu}\right) 
& = & 0 , \\
\partial^{\nu}h_{\mu\nu} - \frac{1}{2}\partial_{\mu}h + \alpha b_{\mu} 
& = & 0 , \\
\mbox{}- m^{2}\partial_{\mu}h 
+ \left(\Box - 2\alpha m^{2}\right) b_{\mu} & = & 0 ,\\
\left(\Box^{2} - 4m^{2}\Box + 6\alpha m^{4}\right) h & = & 0 .
\end{eqnarray}
We also have
\begin{eqnarray}
\left(\Box - 2\alpha m^{2}\right)
\left(\Box^{2} - 4m^{2}\Box + 6\alpha m^{4}\right)
b_{\mu} & = & 0 , \\
\left(\Box - 2\alpha m^{2}\right)
\left(\Box^{2} - 4m^{2}\Box + 6\alpha m^{4}\right)
\left(\partial^{\nu}h_{\mu\nu} - \frac{1}{2}\partial_{\mu}h\right)
& = & 0 , \\
\left(\Box - m^{2}\right)
\left(\Box - 2\alpha m^{2}\right)
\left(\Box^{2} - 4m^{2}\Box + 6\alpha m^{4}\right)
h_{\mu\nu} & = & 0 .
\end{eqnarray}
Two-point functions in this case are
\begin{eqnarray}
\langle h^{\mu\nu}h^{\rho\sigma}\rangle & = & 
\frac{1}{\Box - m^{2}}\left\{
\frac{1}{2}\left(
\eta^{\mu\rho}\eta^{\nu\sigma} 
+ \eta^{\mu\sigma}\eta^{\nu\rho} 
- \eta^{\mu\nu}\eta^{\rho\sigma}\right) \right. \nonumber \\
& & \mbox{}
- \frac{1}{2}(1-2\alpha )\frac{1}{\Box - 2\alpha m^{2}}\left(
\eta^{\mu\rho}\partial^{\nu}\partial^{\sigma}
+ \eta^{\mu\sigma}\partial^{\nu}\partial^{\rho}
+ \eta^{\nu\rho}\partial^{\mu}\partial^{\sigma}
+ \eta^{\nu\sigma}\partial^{\mu}\partial^{\rho}\right) \nonumber \\
& & \mbox{}
- \frac{m^{2}}{\Box^{2} - 4m^{2}\Box + 6\alpha m^{4}}
\left[
\frac{1}{2}\left(\Box - 2\alpha m^{2}\right)
\eta^{\mu\nu}\eta^{\rho\sigma}\right. \nonumber \\
& & 
\makebox[35mm]{}\,
+ ( 1 - 2\alpha )
\left(\eta^{\mu\nu}\partial^{\rho}\partial^{\sigma}
+ \eta^{\rho\sigma}\partial^{\mu}\partial^{\nu}\right) \nonumber \\
& & 
\makebox[34mm]{} \left.\left.\mbox{}
+ 2 ( 1 - 2\alpha )^{2}
\frac{1}{\Box - 2\alpha m^{2}}
\partial^{\mu}\partial^{\nu}\partial^{\rho}\partial^{\sigma}
\right]\right\}\delta , \\
\langle h^{\mu\nu}b^{\rho}\rangle & = & 
\frac{1}{\Box - 2\alpha m^{2}}\left( 
\eta^{\mu\rho}\partial^{\nu} 
+ \eta^{\nu\rho}\partial^{\mu}\right)\delta \nonumber \\
& & \mbox{}
+ \frac{m^{2}}{\Box^{2} - 4m^{2}\Box + 6\alpha m^{4}}
\left[
\eta^{\mu\nu}\partial^{\rho} 
+ 2 ( 1 - 2\alpha )
\frac{1}{\Box - 2\alpha m^{2}}
\partial^{\mu}\partial^{\nu}\partial^{\rho}
\right]\delta , \nonumber \\
& & \\
\langle b^{\mu}b^{\rho}\rangle & = & 
\mbox{}- \frac{2m^{2}}{\Box - 2\alpha m^{2}}
\left[
\eta^{\mu\rho} 
- \frac{\Box - m^{2}}{\Box^{2} - 4m^{2}\Box + 6\alpha m^{4}}
\partial^{\mu}\partial^{\rho}\right]\delta .
\end{eqnarray}
Although these expressins are complicated, 
smooth massless limits are seen to be assured 
for an arbitrary value of the gauge parameter $\alpha$.
For some special values of $\alpha$, 
for example $\alpha = \frac{1}{2}$,
we can have much simpler expressions.

\subsection{ASG model}
In this case the Lagrangian we take is
\begin{equation}
L = \frac{1}{2}h^{\mu\nu}\Lambda_{\mu\nu ,\rho\sigma}h^{\rho\sigma} 
- \frac{m^{2}}{2}\left( h^{\mu\nu}h_{\mu\nu} - \frac{1}{2}h^{2}\right) 
+ b^{\mu}\left(
\partial^{\nu}h_{\mu\nu} - \frac{1}{2}\partial_{\mu}h 
+ \frac{\alpha}{2}b_{\mu}\right) .
\end{equation}
Field equations obtained here are much simpler than in the previous case: 
\begin{eqnarray}
\left(\Box - m^{2}\right) h_{\mu\nu} 
- \frac{1}{2}(1-2\alpha )
\left(\partial_{\mu}b_{\nu} + \partial_{\nu}b_{\mu}\right) 
& = & 0 , \\
\partial^{\nu}h_{\mu\nu} - \frac{1}{2}\partial_{\mu}h + \alpha b_{\mu} 
& = & 0 , \\
\left(\Box - 2\alpha m^{2}\right) b_{\mu} & = & 0 ,
\end{eqnarray}
and further 
\begin{eqnarray}
\left(\Box - 2\alpha m^{2}\right)
\left(\partial^{\nu}h_{\mu\nu} - \frac{1}{2}\partial_{\mu}h\right)
& = & 0 , \\
\left(\Box - m^{2}\right)
\left(\Box - 2\alpha m^{2}\right)
h_{\mu\nu} & = & 0 .
\end{eqnarray}
The structure of two-point functions also becomes simpler as follows:
\begin{eqnarray}
\langle h^{\mu\nu}h^{\rho\sigma}\rangle & = & 
\frac{1}{\Box - m^{2}}\left\{
\frac{1}{2}\left(
\eta^{\mu\rho}\eta^{\nu\sigma} 
+ \eta^{\mu\sigma}\eta^{\nu\rho} 
- \eta^{\mu\nu}\eta^{\rho\sigma}\right) \right. \nonumber \\
& & \makebox[-10mm]{}\left.\mbox{}
- \frac{1}{2}(1-2\alpha )\frac{1}{\Box - 2\alpha m^{2}}\left(
\eta^{\mu\rho}\partial^{\nu}\partial^{\sigma}
+ \eta^{\mu\sigma}\partial^{\nu}\partial^{\rho}
+ \eta^{\nu\rho}\partial^{\mu}\partial^{\sigma}
+ \eta^{\nu\sigma}\partial^{\mu}\partial^{\rho}\right)\right\}\delta ,
\nonumber \\
& & \\
\langle h^{\mu\nu}b^{\rho}\rangle & = & 
\frac{1}{\Box - 2\alpha m^{2}}\left( 
\eta^{\mu\rho}\partial^{\nu} 
+ \eta^{\nu\rho}\partial^{\mu}\right)\delta , \\
\langle b^{\mu}b^{\rho}\rangle & = & 
\mbox{}- \frac{2m^{2}}{\Box - 2\alpha m^{2}}\eta^{\mu\rho}\delta .
\end{eqnarray}
It is seen that Nakanishi's gauge-fixing procedure does work 
for regularizing massless 
singularities in the ASG model too.

\section{BRS Gauge-Fixing Procedure}

\subsection{BRS procedure}
As in the vector case, we start from the usual massive-tensor Lagrangian 
without a gauge-fixing term:
\begin{equation}
L_{h} = \frac{1}{2}h^{\mu\nu}\Lambda_{\mu\nu ,\rho\sigma}h^{\rho\sigma} 
- \frac{m^{2}}{2}\left( h^{\mu\nu}h_{\mu\nu} - ah^{2}\right) ,
\label{eq:501}
\end{equation}
where
\begin{equation}
a = \left\{
\begin{array}{ll}
1           & {\rm for \ the \ FP \ model} , \\
\frac{1}{2} & {\rm for \ the \ ASG \ model} .
\end{array}
\right.
\end{equation}
We introduce a new set of variables $( h_{\mu\nu}', \theta_{\mu} )$ 
to perform a field transformation 
$h_{\mu\nu} \rightarrow ( h_{\mu\nu}', \theta_{\mu} )$ such that
\begin{eqnarray}
h_{\mu\nu} & = & h_{\mu\nu}' 
- \frac{1}{m}\left(\partial_{\mu}\theta_{\nu} 
+ \partial_{\nu}\theta_{\mu}\right) , 
\label{eq:503} \\
\partial^{\nu}h_{\mu\nu}' - \frac{1}{2}\partial_{\mu}h' & = & 0 .
\label{eq:504}
\end{eqnarray}
The Lagrangian (\ref{eq:501}), 
which is independent of the new variables, 
is invariant under the following BRS transformation:
\begin{equation}
\left\{
\begin{array}{lcl}
\delta h_{\mu\nu}' = c_{\mu\nu} , & & 
\delta \bar{c}_{\mu\nu} = ib_{\mu\nu} , \\
\delta \theta_{\mu} = mc_{\mu} ,        & & 
\delta \bar{c}_{\mu} = ib_{\mu} ,
\end{array}
\right.
\end{equation}
where $( c_{\mu\nu}, c_{\mu} )$ and $( \bar{c}_{\mu\nu}, \bar{c}_{\mu} )$ 
denote the FP ghosts and $( b_{\mu\nu}, b_{\mu} )$ indicate the NL fields. 
In order to perform the field transformation (\ref{eq:503}) 
with (\ref{eq:504}), we supplement the Lagrangian (\ref{eq:501}) 
by adding the following BRS gauge-fixing term: 
\begin{eqnarray}
L_{\rm B} & = & 
\mbox{}- i\delta\left[ 
\bar{c}^{\mu\nu}\left(h_{\mu\nu} - h_{\mu\nu}' 
+ \frac{1}{m}\left(
\partial_{\mu}\theta_{\nu} + \partial_{\nu}\theta_{\mu}\right)\right) 
+ \bar{c}^{\mu}\left(\partial^{\nu}h_{\mu\nu}' 
- \frac{1}{2}\partial_{\mu}h' 
+ \frac{\alpha}{2}b_{\mu}\right)\right] 
\nonumber \\
& = & \,
b^{\mu\nu}\left(h_{\mu\nu} - h_{\mu\nu}' 
+ \frac{1}{m}\left(
\partial_{\mu}\theta_{\nu} + \partial_{\nu}\theta_{\mu}\right)\right) 
+ b^{\mu}\left(\partial^{\nu}h_{\mu\nu}' 
- \frac{1}{2}\partial_{\mu}h' 
+ \frac{\alpha}{2}b_{\mu}\right) \nonumber \\
& & \mbox{}
-i\left(\bar{c}^{\mu\nu} + \frac{1}{2}\left(
\partial^{\mu}\bar{c}^{\nu} + \partial^{\nu}\bar{c}^{\mu} 
- \eta^{\mu\nu}\partial_{\rho}\bar{c}^{\rho}\right)\right) 
\left( \frac{}{}c_{\mu\nu} - \left(
\partial_{\mu}c_{\nu} + \partial_{\nu}c_{\mu}\right)\right) 
+ i\bar{c}^{\mu}\Box c_{\mu} .\nonumber \\
& & 
\end{eqnarray}
The path integral is given by
\begin{equation}
Z = \int {\cal D}h_{\mu\nu}{\cal D}h_{\mu\nu}'{\cal D}\theta_{\mu}
{\cal D}b_{\mu\nu}{\cal D}c_{\mu\nu}{\cal D}\bar{c}_{\mu\nu}
{\cal D}b_{\mu}{\cal D}c_{\mu}{\cal D}\bar{c}_{\mu}
\exp i\int d^{4}x\left[L_{h} + L_{\rm B}\right] .
\end{equation}
We integrate out with respect to the variables 
$( b_{\mu\nu}, h_{\mu\nu}, c_{\mu\nu}, \bar{c}_{\mu\nu} )$, 
and then write $h_{\mu\nu}$ over $h_{\mu\nu}'$. 
The result is 
\begin{equation}
Z = \int {\cal D}h_{\mu\nu}{\cal D}\theta_{\mu}{\cal D}b_{\mu}
{\cal D}c_{\mu}{\cal D}\bar{c}_{\mu}
\exp i\int d^{4}x\, L_{\rm T} ,
\end{equation}
where
\begin{eqnarray}
L_{\rm T} & = & 
\frac{1}{2}h^{\mu\nu}\Lambda_{\mu\nu ,\rho\sigma}h^{\rho\sigma} 
\nonumber \\
& & \mbox{}
- \frac{m^{2}}{2}\left[
\left(h_{\mu\nu} - \frac{1}{m}\left(
\partial_{\mu}\theta_{\nu} + \partial_{\nu}\theta_{\mu}
\right)\right)^{2} 
- a\left( h - \frac{2}{m}\partial^{\mu}\theta_{\mu}\right)^{2}
\right] \nonumber \\
& & \mbox{}
+\, b^{\mu}\left(
\partial^{\nu}h_{\mu\nu} - \frac{1}{2}\partial_{\mu}h 
+ \frac{\alpha}{2}b_{\mu}\right) 
+ i\bar{c}^{\mu}\Box c_{\mu} .
\label{eq:509}
\end{eqnarray}
This Lagrangian is invariant under the following BRS transformation:
\begin{equation}
\begin{array}{lclcl}
\delta h_{\mu\nu} = \partial_{\mu}c_{\nu} + \partial_{\nu}c_{\mu} , 
& & 
\delta\theta_{\mu} = mc_{\mu} , 
& & 
\delta\bar{c}_{\mu} = ib_{\mu} . 
\end{array}
\end{equation}
For an Abelian case, which is the case we consider, 
we can neglect the last term in the Lagrangian (\ref{eq:509}) 
because the FP ghosts decouple from the other fields.

\subsection{FP model}
We put $a = 1$ and omit the FP-ghost term in the Lagrangian 
(\ref{eq:509}) to have 
\begin{eqnarray}
L & = & 
\frac{1}{2}h^{\mu\nu}\Lambda_{\mu\nu ,\rho\sigma}h^{\rho\sigma} 
\nonumber \\
& & \mbox{}
- \frac{m^{2}}{2}\left( h^{\mu\nu}h_{\mu\nu} - h^{2}\right) 
- 2m\theta^{\mu}\left(\partial^{\nu}h_{\mu\nu} - \partial_{\mu}h\right) 
- \frac{1}{2}\left(
\partial_{\mu}\theta_{\nu} - \partial_{\nu}\theta_{\mu}\right)^{2}
\nonumber \\
& & \mbox{}
+\, b^{\mu}\left(
\partial^{\nu}h_{\mu\nu} - \frac{1}{2}\partial_{\mu}h 
+ \frac{\alpha}{2}b_{\mu}\right) .
\end{eqnarray}
Field equations derived from this Lagrangian are 
\begin{eqnarray}
\left(\Box - m^{2}\right) h_{\mu\nu} 
+ m\left(\partial_{\mu}\theta_{\nu} + \partial_{\nu}\theta_{\mu}\right) 
\makebox[9mm]{} & & \nonumber \\
\mbox{}- \frac{1}{2}(1-2\alpha )
\left(\partial_{\mu}b_{\nu} + \partial_{\nu}b_{\mu}\right) 
+ \frac{1}{6}\eta_{\mu\nu}\partial^{\rho}b_{\rho}
& = & 0 , \\
\partial^{\nu}h_{\mu\nu} - \frac{1}{2}\partial_{\mu}h + \alpha b_{\mu} 
& = & 0 , \\
\Box h + 2\alpha\partial^{\mu}b_{\mu} 
& = & 0 , \\
\Box\theta_{\mu} + \alpha mb_{\mu} 
- \frac{1}{6m}\partial_{\mu}\partial^{\nu}b_{\nu} 
& = & 0 , \\
\Box b_{\mu} & = & 0 .
\end{eqnarray}
We also have 
\begin{eqnarray}
\Box\left(\partial^{\nu}h_{\mu\nu} - \frac{1}{2}\partial_{\mu}h\right)
& = & 0 , \\
\Box^{2}h & = & 0 ,\\
\Box^{2}\theta_{\mu} & = & 0 ,\\
\Box^{2}\left(\Box - m^{2}\right) h_{\mu\nu} & = & 0 .
\end{eqnarray}
Two-point functions obtained are the following:
\begin{eqnarray}
\langle h^{\mu\nu}h^{\rho\sigma}\rangle & = & 
\frac{1}{\Box - m^{2}}\left\{
\frac{1}{2}\left(
\eta^{\mu\rho}\eta^{\nu\sigma} 
+ \eta^{\mu\sigma}\eta^{\nu\rho} 
- \eta^{\mu\nu}\eta^{\rho\sigma}\right) \right. \nonumber \\
& & \mbox{}
- \frac{1}{2}\left[
(1-2\alpha )\frac{1}{\Box} 
+ 2\alpha\frac{m^{2}}{\Box^{2}}\right]\left(
\eta^{\mu\rho}\partial^{\nu}\partial^{\sigma}
+ \eta^{\mu\sigma}\partial^{\nu}\partial^{\rho}
+ \eta^{\nu\rho}\partial^{\mu}\partial^{\sigma}
+ \eta^{\nu\sigma}\partial^{\mu}\partial^{\rho}\right) \nonumber \\
& & \makebox[-1mm]{}\left.\mbox{}
+ \frac{2}{3}\left(
\frac{1}{2}\eta^{\mu\nu} 
+ \frac{\partial^{\mu}\partial^{\nu}}{\Box}\right)\left(
\frac{1}{2}\eta^{\rho\sigma} 
+ \frac{\partial^{\rho}\partial^{\sigma}}{\Box}\right)
\right\} \delta ,\\ 
\langle h^{\mu\nu}b^{\rho}\rangle & = & 
\frac{1}{\Box}\left( 
\eta^{\mu\rho}\partial^{\nu} 
+ \eta^{\nu\rho}\partial^{\mu}\right)\delta , \\
\langle h^{\mu\nu}\theta^{\rho}\rangle & = & 
\left\{
\frac{1}{6m}\frac{1}{\Box}\eta^{\mu\nu}\partial^{\rho} 
- \alpha m\frac{1}{\Box^{2}}\left(
\eta^{\mu\rho}\partial^{\nu} + \eta^{\nu\rho}\partial^{\mu}\right) 
+ \frac{1}{3m}\frac{1}{\Box^{2}}
\partial^{\mu}\partial^{\nu}\partial^{\rho}
\right\}\delta , \label{eq:523} \\
\langle b^{\mu}b^{\rho}\rangle & = & 0 , \\
\langle b^{\mu}\theta^{\rho}\rangle & = & 
\frac{m}{\Box}\eta^{\mu\rho}\delta , \\
\langle\theta^{\mu}\theta^{\rho}\rangle & = & 
\left\{
\frac{1}{2}\frac{1}{\Box}\left(
1 - 2\alpha\frac{m^{2}}{\Box}\right)\eta^{\mu\rho} 
- \frac{1}{6m^{2}}\frac{1}{\Box}\left(
1 - \frac{m^{2}}{\Box}\right)
\partial^{\mu}\partial^{\rho}\right\}\delta .
\label{eq:526}
\end{eqnarray}
We see there still remain massless singularities. 
The singularities found in (\ref{eq:315}) have been driven away 
indeed. 
However, the new singularities, though weaker than the 
original ones, have appeared in the $\theta$-sector 
(\ref{eq:523}) and (\ref{eq:526}).
It follows that the BRS gauge-fixing procedure 
cannot drive away all the massless singularities of the FP model
although this procedure does reduce the degree of singularities.

\subsection{ASG model}
In this case we set $a = \frac{1}{2}$ in the Lagrangian (\ref{eq:509}). 
Neglecting the FP-ghost term again, we have 
\begin{eqnarray}
L & = & 
\frac{1}{2}h^{\mu\nu}\Lambda_{\mu\nu ,\rho\sigma}h^{\rho\sigma} 
\nonumber \\
& & \mbox{}
- \frac{m^{2}}{2}\left( h^{\mu\nu}h_{\mu\nu} - \frac{1}{2}h^{2}\right) 
- 2m\theta^{\mu}\left(
\partial^{\nu}h_{\mu\nu} - \frac{1}{2}\partial_{\mu}h\right) 
- \partial_{\mu}\theta_{\nu}\partial^{\mu}\theta^{\nu} \nonumber \\
& & \mbox{}
+ b^{\mu}\left(
\partial^{\nu}h_{\mu\nu} - \frac{1}{2}\partial_{\mu}h 
+ \frac{\alpha}{2}b_{\mu}\right) .
\end{eqnarray}
Field equations in this case are 
\begin{eqnarray}
\left(\Box - m^{2}\right) h_{\mu\nu} 
+ m\left(\partial_{\mu}\theta_{\nu} + \partial_{\nu}\theta_{\mu}\right) 
& & \nonumber \\
\mbox{}- \frac{1}{2}(1-2\alpha )
\left(\partial_{\mu}b_{\nu} + \partial_{\nu}b_{\mu}\right) 
& = & 0 , \\
\partial^{\nu}h_{\mu\nu} - \frac{1}{2}\partial_{\mu}h + \alpha b_{\mu} 
& = & 0 , \\
\Box\theta_{\mu} + \alpha mb_{\mu} & = & 0 , \\
\Box b_{\mu} & = & 0 .
\end{eqnarray}
Further we have 
\begin{eqnarray}
\Box\left(\partial^{\nu}h_{\mu\nu} - \frac{1}{2}\partial_{\mu}h\right)
& = & 0 , \\
\Box^{2}\theta_{\mu} & = & 0 ,\\
\Box^{2}\left(\Box - m^{2}\right) h_{\mu\nu} & = & 0 .
\end{eqnarray}
Two-point functions are calculated as 
\begin{eqnarray}
\langle h^{\mu\nu}h^{\rho\sigma}\rangle & = & 
\frac{1}{\Box - m^{2}}\left\{
\frac{1}{2}\left(
\eta^{\mu\rho}\eta^{\nu\sigma} 
+ \eta^{\mu\sigma}\eta^{\nu\rho} 
- \eta^{\mu\nu}\eta^{\rho\sigma}\right) \right. \nonumber \\
& & \makebox[-15mm]{}\left.\mbox{}
- \frac{1}{2}\left[
(1-2\alpha )\frac{1}{\Box} 
+ 2\alpha\frac{m^{2}}{\Box^{2}}\right]\left(
\eta^{\mu\rho}\partial^{\nu}\partial^{\sigma}
+ \eta^{\mu\sigma}\partial^{\nu}\partial^{\rho}
+ \eta^{\nu\rho}\partial^{\mu}\partial^{\sigma}
+ \eta^{\nu\sigma}\partial^{\mu}\partial^{\rho}\right)\right\}\delta , 
\nonumber \\ 
& & \\
\langle h^{\mu\nu}b^{\rho}\rangle & = & 
\frac{1}{\Box}\left( 
\eta^{\mu\rho}\partial^{\nu} 
+ \eta^{\nu\rho}\partial^{\mu}\right)\delta , \\
\langle h^{\mu\nu}\theta^{\rho}\rangle & = & 
\mbox{}- \alpha m\frac{1}{\Box^{2}}\left(
\eta^{\mu\rho}\partial^{\nu} + \eta^{\nu\rho}\partial^{\mu}\right) 
\delta , \\
\langle b^{\mu}b^{\rho}\rangle & = & 0 , \\
\langle b^{\mu}\theta^{\rho}\rangle & = & 
\frac{m}{\Box}\eta^{\mu\rho}\delta , \\
\langle\theta^{\mu}\theta^{\rho}\rangle & = & 
\frac{1}{2}\frac{1}{\Box}\left(
1 - 2\alpha\frac{m^{2}}{\Box}\right)\eta^{\mu\rho}\delta .
\end{eqnarray}
Compare these expressions with the corresponding ones (\ref{eq:319a}) 
for the original ASG model. 
It is seen that 
the BRS gauge-fixing procedure have been able to regularize the 
massless singularities involved in the ASG model.

\section{Summary and Discussion}
In this paper we have studied how to construct massive tensor theories 
with smooth massless limits. 
We have taken up the FP and ASG models for a linearized massive tensor 
field, 
and applied Nakanishi's and the BRS gauge-fixing procedures 
to each model. 
It has been found that the ASG model can be regularized by both of 
the procedures, 
while the FP model only by Nakanishi's procedure.
We have thus obtained three kinds of regularized massive tensor 
theories without massless singularities:
N-regularized FP model, N-regularized ASG model 
and BRS-regularized ASG model.

In order to construct a complete nonlinear theory, 
it is desirable to have linearized theories with higher symmetry 
properties. 
BRS symmetry seems to play an essential role in this respect. 
The BRS procedure just provides this symmetry, 
but Nakanishi's procedure does not. 
It follows that the BRS-regularized ASG model may be most promising 
for our purpose. 
Detailed discussions along this line will be made in a future 
publication.

\section*{Acknowledgements}
We would like to thank M.~Hirayama and H.~Yamakoshi for discussions.




\begin{thebibliography}{99}
\bibitem{rf:1}
H.~van Dam and M.~Veltman, \NP{B22,1970,397}.
\bibitem{rf:2}
T.~Kimura, \PTP{55,1976,1259}.
\bibitem{rf:3}
C.~Fronsdal and W.F.~Heidenreich, \ANN{215,1992,51}.
\bibitem{rf:4}
N.~Nakanishi, \JL{Prog.~Theor.~Phys.~Suppl.,51,1972,1}.
\bibitem{rf:5}
K.-I.~Izawa, \PTP{88,1992,759}.
\end{thebibliography}
\end{document}